\documentclass[12pt]{article}

\usepackage{epsf}
\usepackage[dvips]{graphicx}
\usepackage{epic}
\usepackage{eepic}

\begin{document}
%%%%%%%%%%%%%%%%%%%%%%%%%%%%%%%%%%%%%%%%%%%%%%%%%%%%%%%%%%%
\def\bc{\begin{center}}
\def\ec{\end{center}}
\def\beq{\begin{equation}}
\def\eeq{\end{equation}}
\def\noi{\noindent}
\def\ol#1{\overline{#1}}

\def\at#1{\left. \right|^{}_{#1}}
\def\hs#1{\hspace*{#1cm}}
\def\la{\langle}
\def\ra{\rangle}

\def\av#1{\langle #1 \rangle}
\def\ave#1{\langle {#1} \rangle}
\def\avr#1#2{\langle {#1} \rangle^{}_{#2}}
\def\avA#1{\langle #1 \rangle_A^{}}
\def\avB#1{\langle #1 \rangle_B^{}}
\def\avAB#1{\avA {\avB {{#1}}}}
\def\avt#1{\avA {\avB {\overline{#1}}}}

\def\pra#1{\prod_{{#1}=1}^A}
\def\prda#1{\prod_{{#1}=1}^A da_{#1}}
\def\prhda#1{\prod_{{#1}=1}^A \hat{d}a_{#1}}
\def\prTda#1{\prod_{{#1}=1}^A T_A(a_{#1}) da_{#1}}

\def\prb#1{\prod_{{#1}=1}^B}
\def\prdb#1{\prod_{{#1}=1}^B db_{#1}}
\def\prhdb#1{\prod_{{#1}=1}^B \hat{d}b_{#1}}
\def\prTdb#1{\prod_{{#1}=1}^B T_B(b_{#1}) db_{#1}}

\def\sumA#1{\sum_{{#1}=1}^A}
\def\sumB#1{\sum_{{#1}=1}^B}

\def\vybk{\{k_1,...,k_n\}}
\def\vybkd{\{k_{n+1},...,k_B\}}
\def\QQ{Q^{(12)}}
\def\ss{\sigma^{(j_1 j_2)}}
\def\ssi{\sigma^{(12)}}
\def\sig#1{\sigma_{#1}}
\def\Xj#1{X_{j_{#1}}}
\def\pj#1{p_{j_{#1}}}
\def\qj#1{q_{j_{#1}}}
\def\Cai{\{a_i\}}
\def\Cbk{\{b_k\}}
\def\hda#1{\hat{d}a_{#1}}
\def\hdb#1{\hat{d}b_{#1}}
%%%%%%%%%%%%%%%%%%%%%%%%%%%%%%%%%%%%%%%%%%%%%%%%%%%%%%%%%%%

\begin{center}
{\bfseries
ON PARTICIPANTS NUMBER FLUCTUATIONS\\
FOR GIVEN CENTRALITY AA-INTERACTIONS\\
IN THE CLASSICAL GLAUBER APPROACH}

\vskip 5mm

V.V. Vechernin$^{\dag}$

\vskip 5mm

{\small
{\it
V.A.Fock Institute of Physics,
St.-Petersburg State University
}
\\
$\dag$ {\it
E-mail: vechernin@pobox.spbu.ru
}}
\end{center}

\vskip 5mm

\begin{center}
\begin{minipage}{150mm}
\centerline{\bf Abstract}
In the framework of the classical Glauber approach
the exact analytical expression for the variance of the number of participants
(wounded nucleons)
for given centrality AA interactions
is presented.
It's shown, that
in the case of nucleus-nucleus collisions
along with the optical approximation term
the additional "contact" term
appears.
%arises.

The numerical calculations
for PbPb collisions at SPS energies
show that at intermediate values of the impact parameter
the "optical" and "contact" terms contributions
to the variance of the participants number
are of the same order and
their sum is in a good agreement
with the results of independent MC simulations
of this process.

The correlation between the numbers of participants
in colliding nuclei is taken into account.
In particular
it's demonstrated
that in PbPb collisions at SPS energies
the variance of the total number of participants
approximately three times exceeds the Poisson one
in the impact parameter region 10-12\,Fm.
The fluctuations of the number of collisions are also discussed.
\end{minipage}
\end{center}

\vskip 10mm

%%%% figQ %%%%%%%%%%%%%%%%%%%%%%%%%%%%%%%%%%%%%%%%%%%%%%%%%%%%
\def\figQ{\begin{figure}[b]

\unitlength=0.8mm
%\linethickness{0.5 pt}
\thicklines
\bc
%==================================
\begin{picture}(170,30)(-20,0)
%\filltype{black}
%\blacken
\drawline[-20](10,5)(10,30)
\drawline[-20](20,0)(20,30)
\drawline[-20](30,5)(30,25)
\drawline[-20](40,0)(40,25)
 \put(10,5){\circle*{1}}
 \put(10,30){\circle*{1}}
 \put(20,0){\circle*{1}}
 \put(20,30){\circle*{1}}
 \put(30,5){\circle*{1}}
 \put(30,25){\circle*{1}}
 \put(40,0){\circle*{1}}
 \put(40,25){\circle*{1}}
\put(0,0){\line(1,0){50}}
\put(0,5){\line(1,0){50}}
\put(0,25){\line(1,0){50}}
\put(0,30){\line(1,0){50}}
\put(-1,0){\makebox(0,0)[rc]{$2'$}}
\put(-1,5){\makebox(0,0)[rc]{$1'$}}
\put(-1,25){\makebox(0,0)[rc]{$2$}}
\put(-1,30){\makebox(0,0)[rc]{$1$}}
 \put(80,15) {\makebox(0,0)[cc]{\Large{$\bullet$}}}
 \put(95,30) {\makebox(0,0)[cc]{\rule{2mm}{2mm}}}
 \put(95,0)  {\makebox(0,0)[cc]{\rule{2mm}{2mm}}}
 \put(110,15){\makebox(0,0)[cc]{\Large{$\bullet$}}}
\put(78,15) {\makebox(0,0)[rc]{$1'$}}
\put(112,15){\makebox(0,0)[lc]{$2'$}}
\put(95,32) {\makebox(0,0)[cb]{$1$}}
\put(95,-2)  {\makebox(0,0)[ct]{$2$}}
\drawline[-20](80,15)(95,30)
\drawline[-20](95,30)(110,15)
\drawline[-20](110,15)(95,0)
\drawline[-20](95,0)(80,15)

\end{picture}
%==================================
\ec
\caption[dummy]{\label{loop}
An example of the loop diagram in quantum Glauber approach
to AB-collisions.  1 and 2 - nucleons of the nucleus A;\ \
$1'$ and $2'$ - nucleons of the nucleus B
(see \cite{Boreskov88,Pak79,Braun878890} for details).
}
\end{figure}}
%%%% end figQ %%%%%%%%%%%%%%%%%%%%%%%%%%%%%%%%%%%%%%%%%%%%%%%%%%%%
% END Figures  XXXXXXXXXXXXXXXXXXXXXXXXXXXXXXXXXXXXXXXXXXXXXXXXXXXXXX

\section{Variance of participants number in one nucleus}

At first we consider the variance $V[N^A_w(b)]$ of the number of participants (wounded nucleons)
in one of the colliding nuclei $N^A_w(b)$
at a fixed value of the impact parameter $b$.
In the framework of pure classical, probabilistic approach to nucleus-nucleus collisions,
formulated in \cite{Bialas76}, we find for the mean value and for the variance of $N^A_w(b)$:
\begin{equation}
\langle N^A_w(b) \rangle = A P(b) \ ,
\label{mean}
\end{equation}
\begin{equation}
V[N^A_w(b)] =AP(b)Q(b)+A(A-1)[Q^{(12)}(b)-Q^2(b)] \ ,
\label{disp}
\end{equation}
where $P(b)=1-Q(b)$. For $Q(b)$ and $Q^{(12)}(b)$ we have:
$$
Q(b)=\int da_1 T_A(a_1) [1-\sigma_1(a_1)]^B \ ,
$$
\begin{equation}
Q^{(12)}(b)= \int da_1 da_2 T_A(a_1) T_A(a_2) [1-\sigma_{1}(a_1)-\sigma_{1}(a_2)+\sigma^{(12)}(a_1,a_2)]^B \ ,
\label{QQ}
\end{equation}
where
$$
\sigma_{1}(a_{1,2}) \equiv \int db_1 T_B(b_1) \sigma(a_{1,2}-b_1+b)\ ,
%\hspace*{1cm}
%\sigma_{2} \equiv \int db_1 T_B(b_1) \sigma(a_2-b_1+b) \ ,
$$
\begin{equation}
\sigma^{(12)}(a_1,a_2)\equiv \int db_1 T_B(b_1) \sigma(a_{1}-b_1+b)\sigma(a_{2}-b_1+b) \ .
\label{ssi}
\end{equation}
Here $T_A$ and $T_B$ are the profile functions of the colliding nuclei $A$ and $B$;
the $\sigma (a)$ is the probability of inelastic interaction of two nucleons
at the impact parameter value~$a$  ($\int \sigma (a) da = \sigma_{NN}^{in}\equiv\sigma$)
and all integrations imply the integration over
two-dimensional vectors in the impact parameter plane.

The formula (\ref{mean}) and the first term in (\ref{disp}) correspond to the naive picture
(so-called optical approximation)
implying that in the case of $AB$-collision at the impact parameter $b$
one can use the binomial distribution (\ref{opt}) for $N^A_w(b)$
with some averaged probability $P(b)$ of inelastic interaction
of a nucleon of the nucleus $A$ with nucleons of the nucleus $B$.
At that the $P(b)$ is considered the same for all
nucleons of the nucleus $A$.

The whole expression (\ref{disp})
is the result of more accurate calculation
(see appendix \ref{ap:A}),
when one uses probabilistic considerations
taking into account
the impact parameter plane positions
of nucleons in the nuclei $A$ and $B$
and averaging then over these positions:
\begin{equation}
V[N^A_w(b)] =\av{{N_w^A(b)}^2}-\av{N^A_w(b)}^2 \ ,
\label{defvar}
\end{equation}
where
\begin{equation}
\av X \equiv \avt X \equiv
\int \ol{X} \prTdb{k} \prTda{i} \ .
\label{defavr}
\end{equation}
Here $\ol{X}$ means average of some variate $X$ at fixed positions of
all nucleons in $A$ and $B$; $\avr{\ }{A}$ and $\avr{\ }{B}$
mean averaging over positions of these nucleons.

In the limit $r_N\ll R_A, R_B$ formulae (\ref{ssi}) reduce to
$$
\sig{1}(a_{1,2}) \approx  \sigma\,T_B(a_{1,2}+b) \ ,
$$
\beq
\ssi(a_1,a_2) \approx I(a_{2}-a_{1}) \cdot T_B(a_{1}+b) \ , \ {\rm with}
\hspace*{5mm} I(a)\equiv \int ds \, \sigma(s)\, \sigma(s+a) \ .
\label{ssiAPPROX}
\eeq
Note that in this limit
the $Q(b)$ and hence the first term of (\ref{disp}) and (\ref{mean})
depend only on the integral inelastic $NN$ cross-section $\sigma\equiv\sigma_{NN}^{in}$,
but the $Q^{(12)}(b)$ entering the second term of (\ref{disp})
depends also on the shape of the function $\sigma (b)$
through the integral $I(a)$ (see equation (\ref{ssiAPPROX})).

%%%%%%%%%%%%%%%%%%%%%
Note also that using of the approximation $\sigma (b)=\sigma \delta(b)$
for $NN$ interaction gives the same result (as taking of the limit $r_N\ll R_A, R_B$)
only for naive part of the answer,
which is expressed through $Q(b)$.
If one will try to use the approximation $\sigma (b)=\sigma \delta(b)$
for to calculate $\QQ(b)$,
then one gets $I(a)=\sigma^2 \delta(a)$
and $\ssi=\sigma^2 \delta(a_{2}-a_{1}) \cdot T_B(a_{1}+b)$,
which leads to infinite $\QQ(b)$ at $B\geq 2$.
Meanwhile, for any correct approximation of $\sigma (b)$,
when $\sigma (b)\leq 1$ in correspondence with its probabilistic interpretation,
we find definite finite answer for $\QQ(b)$.

In our numerical calculations we'll use for $NN$ interaction
the "black disk" approximation:
\beq
\sigma (b)= \theta(r_N-|b|)\ ,
\label{black}
\eeq
or Gauss approximation:
\beq
\sigma (b)=\exp(-b^2/r^2_N)\ .
\label{Gauss}
\eeq
In both cases $\sigma=\pi r^2_N$.
For the nuclear profile functions $T_A$ and $T_B$
we'll use the standard Woods-Saxon approximation.

%%%%%%%%%%%%%%%%%%%%%%%%%%%%%
We would like to emphasize that nontrivial second term in (\ref{disp})
arises only in the case of nucleus-nucleus collisions.
For $A=1$ or $B=1$ it's equal to zero.
At $A=1$ due to explicit factor $A-1$ in (\ref{disp})
and at $B=1$ due to fact that in this case $\QQ(b)=Q^2(b)$.
This corresponds to the well known fact that for nucleus-nucleus collisions
the Glauber approach doesn't reduce to the so-called optical approximation
even in the limit $r_N\ll R_A, R_B$ (see, for example, \cite{Boreskov88}).

This additional term,
which arises in (\ref{disp}) in the case of nucleus-nucleus collisions,
depends, as we have mentioned, not only on integral value
of inelastic $NN$ cross-section
$\sigma_{NN}^{in}\equiv\sigma=\int \sigma (a) da$,
but also on the shape of the function $\sigma (b)$,
i.e. on the details of $NN$ interaction at $r_N$ distances,
which are much smaller than the typical nuclear distances.
In quantum Glauber approach this corresponds to the fact that
in the case of $AA$ collisions,
in contrast with $pA$ collisions,
the loop diagrams of the type shown in Fig.\ref{loop} appear.
The typical momentum corresponding this loop integration is
much larger than the typical nuclear momenta
(which corresponds to smaller than typical nuclear distances)
and one encounters the "contact" terms problem
(see, for example, \cite{Boreskov88,Pak79,Braun878890}).
The second term in formula (\ref{disp}) is the manifestation of
this problem at the classical level.

\begin{figure}[t]
\centerline{
\includegraphics[width=90mm,angle=-90]{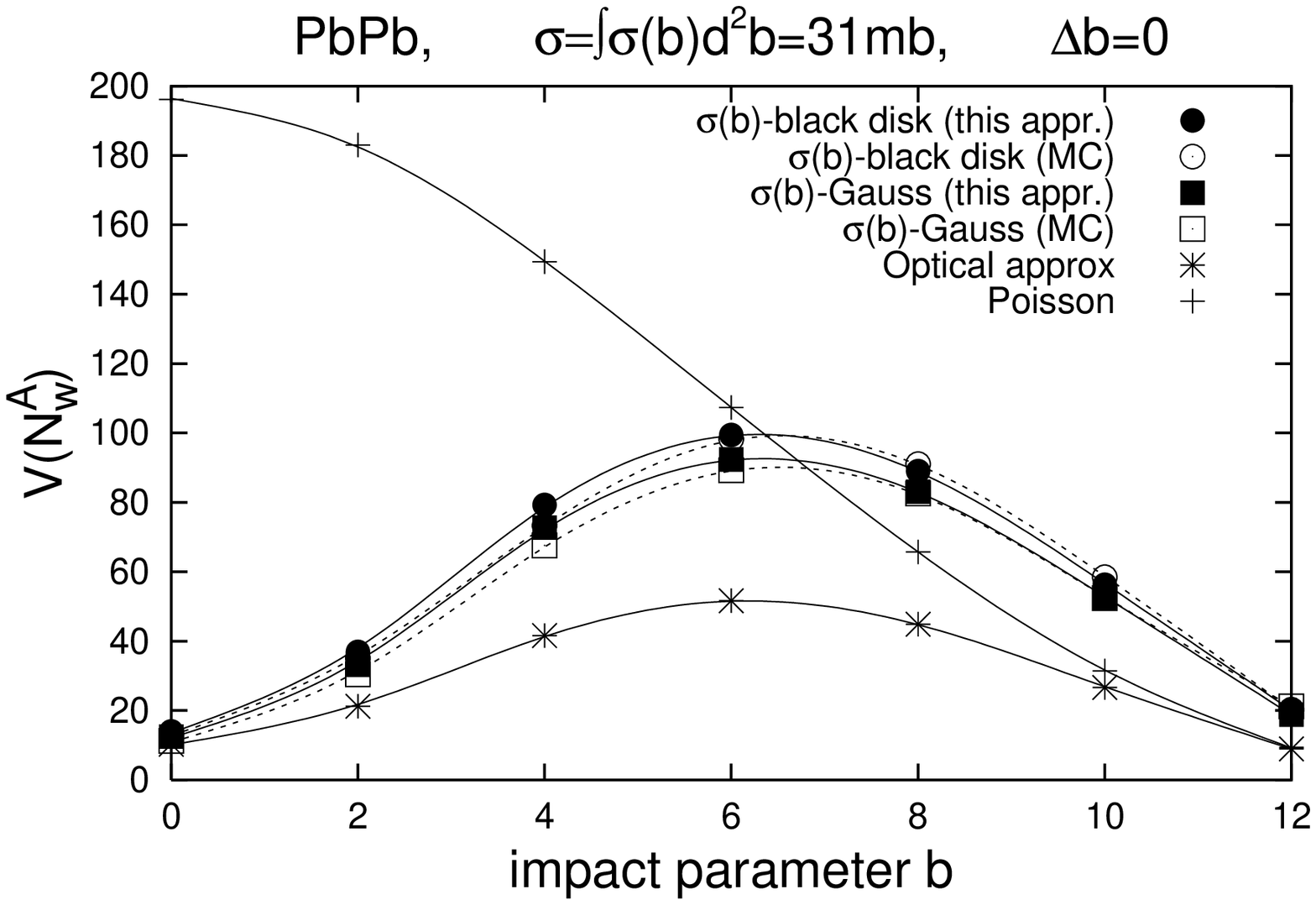}
}
\caption[dummy]{\label{vnwpoia}
The variance of the number of wounded nucleons in one nucleus
for PbPb collisions at SPS energies
($\sigma\equiv\sigma_{NN}^{in}$=31\,mb, $r_N$=1\,Fm)
as a function of the impact parameter $b$.\
The points {\Large $\bullet$} and $\rule[0.2mm]{2.3mm}{2.3mm}$
- results of numerical calculations
by formulae (\ref{disp}), (\ref{QQ}) and (\ref{ssiAPPROX})
using respectively the "black disk" (\ref{black})
and Gaussian (\ref{Gauss}) approximations for $NN$ interaction; \
{\Large $\circ$} and \raisebox{1.5mm}{\framebox[2mm][c]{}}
 - results of independent MC calculations
using for $NN$ interaction
"black disk" (\ref{black}) or Gaussian (\ref{Gauss}) approximation; \
{\Large \raisebox{0mm}{${\ast}$}}
- "optical" approximation (the first term in formula (\ref{disp})); \
\raisebox{0mm}{+}
 - the Poisson variance: $V[N^A_w(b)]=\av{N^A_w(b)}$.
The curves are shown to guide eyes.
}
\end{figure}

\figQ

The numerical evaluation of the contribution of
the additional - "contact" term in (\ref{disp}) are presented
in Fig.\ref{vnwpoia} for PbPb collisions at SPS energies
($\sigma\equiv\sigma_{NN}^{in}$=31\,mb, $r_N$=1\,Fm).
For the control we also carried out independent Monte-Carlo calculations
of the mean values and the variances involved presenting the results on the same
figures.
All calculations were done at fixed values of the impact parameter $b$ ($\Delta b=0$).

In Fig.\ref{vnwpoia} we see that
the calculated "contact" term in (\ref{disp})
is essential and
gives approximately the same contribution to the $N^A_w(b)$ variance
for PbPb collisions at intermediate values of $b$,
as the first "optical" term in (\ref{disp}).
We see also that the results of independent MC calculations of the $N^A_w(b)$ variance
are in a good agreement with the results of analytical calculations by formula (\ref{disp}),
but only with taking into account its second term.

Note that due to this "contact" term the $N^A_w(b)$ variance
is larger than the Poisson one
for peripheral PbPb collisions (at $b>7$\,Fm).
The week dependence of the results on details
of $NN$ interaction at nucleon distances is also seen.
The results
%for the $N^A_w(b)$ variance
lay systematically slightly higher
in the case of using the "black disk" (\ref{black}) approximation
for $\sigma (b)$, than in the case of using the Gaussian (\ref{Gauss}) approximation
with the same value of $\sigma$.

\begin{figure}[t]
\centerline{
\includegraphics[width=90mm,angle=-90]{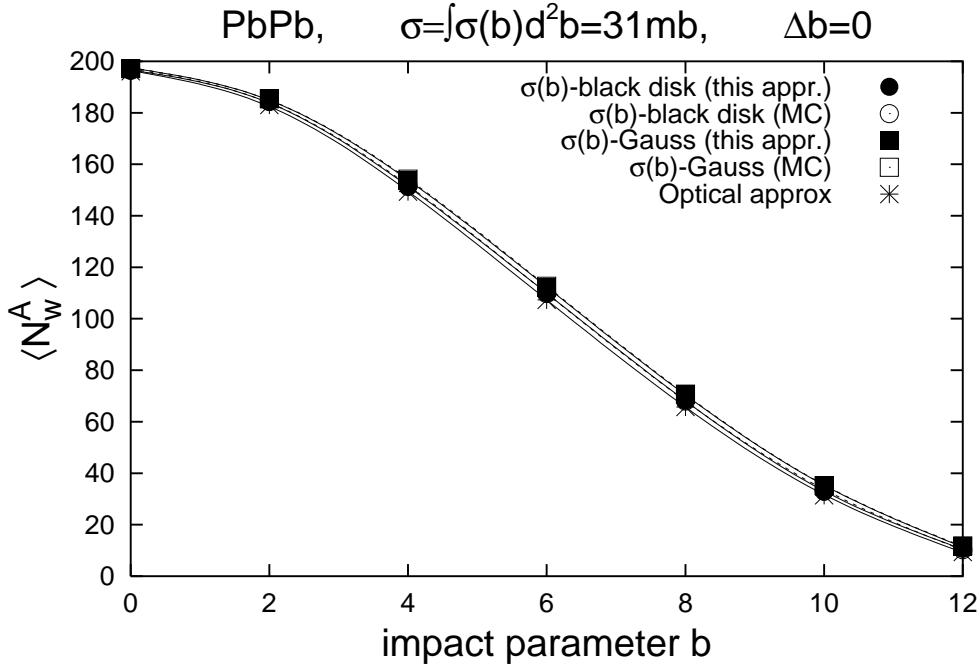}
}
\caption[dummy]{\label{nwa}
The same as in Fig.\ref{vnwpoia}, but for
the mean number of wounded nucleons in one nucleus,
calculated by formulae (\ref{mean}), (\ref{QQ}) and (\ref{ssi});
{\Large \raisebox{0mm}{${\ast}$}}
- "optical" approximation, calculated using formulae (\ref{mean}), (\ref{QQ}) and (\ref{ssiAPPROX}).
}
\end{figure}

For the mean value $\av{N^A_w(b)}$, in contrast to the variance of $N^A_w(b)$,
the exact answer
coincides with the "optical" approximation result
(see formula (\ref{mean}) and appendix \ref{ap:A})
and depends only on $\sigma$.
%and doesn't depend on the type of $\sigma(b)$,
In Fig.\ref{nwa} we see that MC calculations also confirm this result.

\section{Variance of the total number of participants}

\begin{figure}[t]
\centerline{
\includegraphics[width=90mm,angle=-90]{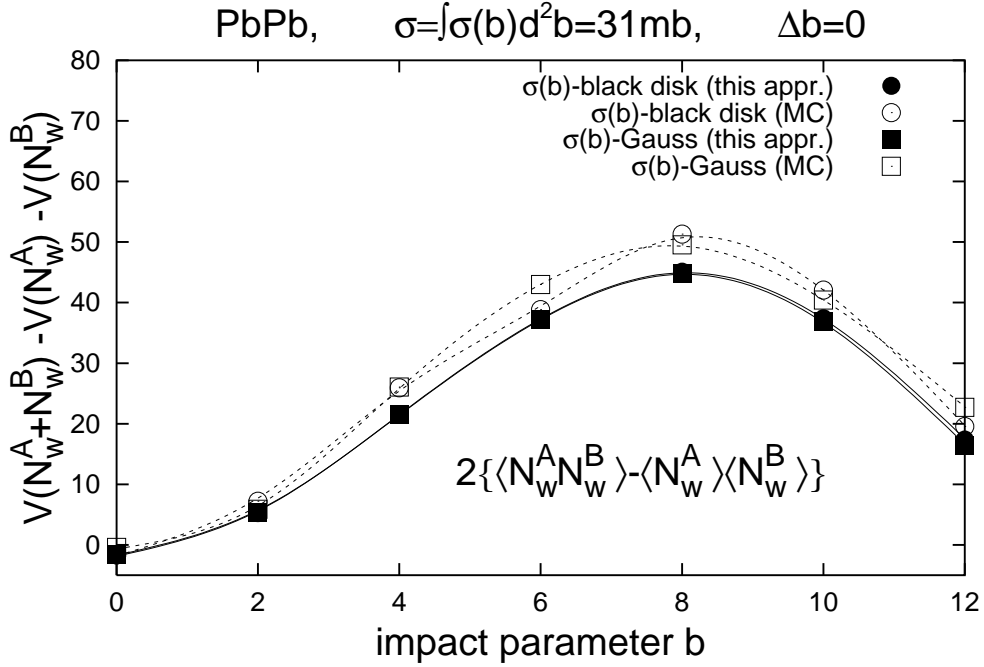}
}
\caption[dummy]{\label{vnw11ab}
The correlator between the numbers
of wounded nucleons in colliding nuclei,
calculated by formulae (\ref{corr})-(\ref{tildesigma})
and by independent MC simulations.
The notations are the same as in Fig.\ref{vnwpoia}.
}
\end{figure}

Now we pass to the calculation of the variance of the total number of participants
$V[N^A_w(b)+N^B_w(b)]$ at a fixed value of the impact parameter $b$.
Clear, that we simply have for the mean value
\beq
\av{N^A_w(b)+N^B_w(b)}= \av{N^A_w(b)} +\av{N^B_w(b)}
\label{meantot}
\eeq
and by (\ref{defvar}) for the variance
\beq
V[N^A_w(b)+N^B_w(b)] = V[N^A_w(b)]+V[N^B_w(b)]
+2\{ \av{N_w^A(b)N^B_w(b)}-\av{N^A_w(b)}\av{N^B_w(b)}\} \ .
\label{vartot}
\eeq
In naive approach ("optical approximation") there are no correlations:
$$
\av{N_w^A(b)N^B_w(b)}-\av{N^A_w(b)}\av{N^B_w(b)}=0 \ .
$$
More accurate calculations (see appendix \ref{ap:B}),
based on formulae (\ref{defvar}) and (\ref{defavr}),
lead to
\begin{equation}
\av{N_w^A(b)N^B_w(b)}-\av{N^A_w(b)}\av{N^B_w(b)}=
AB[Q^{(11)}(b)-Q(b)\widetilde Q(b)] \ ,
\label{corr}
\end{equation}
where
$$
Q(b)=\int da_1 T_A(a_1) [1-\sigma_1(a_1)]^B \ ,
\hs{1}
\widetilde Q(b)=\int db_1 T_B(b_1) [1-\widetilde\sig{1}(b_1)]^A \ ,
$$
\begin{equation}
Q^{(11)}(b)= \int da_1 db_1 T_A(a_1) T_B(b_1) [1-\sigma_{1}(a_1)]^{B-1}
[1-\widetilde\sigma_{1}(b_1)]^{A-1}[1-\sigma(a_1-b_1+b)] \ .
\label{tildeQ}
\end{equation}
Here
$$
\sigma_{1}(a_{1}) \equiv \int db_1 T_B(b_1) \sigma(a_{1}-b_1+b)
\approx  \sigma\,T_B(a_{1}+b) \ ,
$$
\begin{equation}
\widetilde\sigma_{1}(b_{1}) \equiv \int da_1 T_A(a_1) \sigma(a_1-b_1+b)
\approx \sigma T_A(b-b_1)
\label{tildesigma}
\end{equation}
Note that $Q(b)$ and $\sigma_{1}(a_{1})$ are the same as in formulae
(\ref{QQ}), (\ref{ssi}) and (\ref{ssiAPPROX}).

The results of numerical calculations of
the correlator (\ref{corr})
by formulae (\ref{tildeQ}) and (\ref{tildesigma})
together with the results of independent Monte-Carlo calculations
for PbPb collisions at SPS energies
are presented in Fig.\ref{vnw11ab}.

Comparing Fig.\ref{vnw11ab} with Fig.\ref{vnwpoia} we see that
the contribution of this correlator to the
variance of the total number of participants
at intermediate values of $b$
is about half of the variance for one nucleus $V[N^A_w(b)]$
and is about the contribution of first "optical" term in (\ref{disp}).
(Note that the relative contribution of the correlator
to $V[N^A_w(b)+N^B_w(b)]$ even greater at large values of $b$,
$b\geq10$.)
%At that t
The results are again in a good agreement with the results of MC calculations.
%$\av{N_w^A(b)N^B_w(b)}-\av{N^A_w(b)}\av{N^B_w(b)}$.

\begin{figure}[t]
\centerline{
\includegraphics[width=90mm,angle=-90]{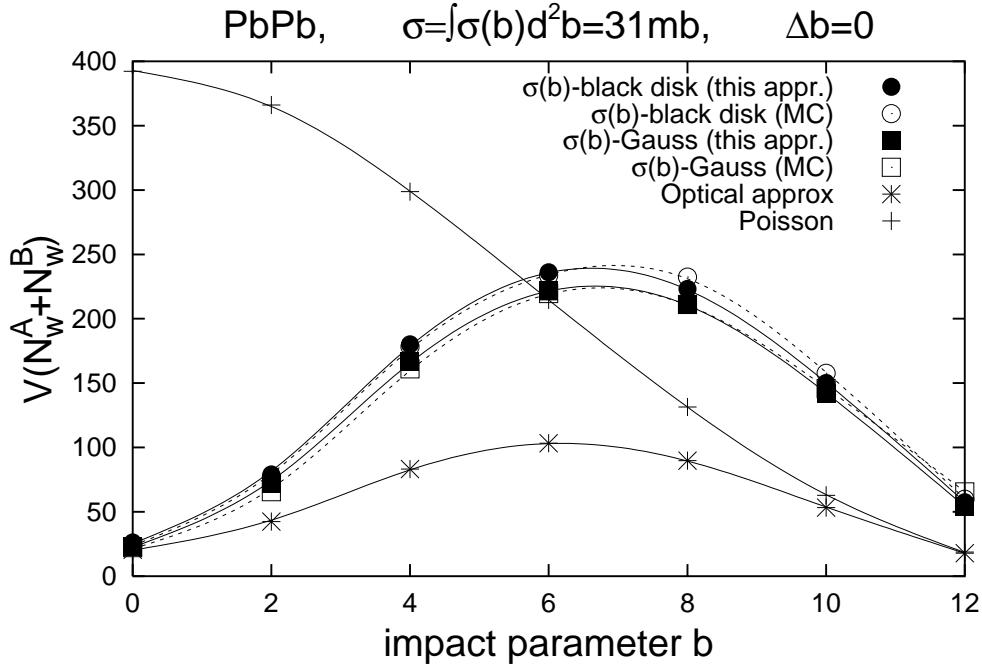}
}
\caption[dummy]{\label{vnwpoiab}
The same as in Fig.\ref{vnwpoia}, but for
the variance of the total number of wounded nucleons
$N_w(b)\equiv N^A_w(b)+N^B_w(b)$ in colliding nuclei;
the variance $V[N_w(b)]$ is calculated by formulae
(\ref{disp}), (\ref{QQ}), (\ref{ssiAPPROX}) and  (\ref{vartot})-(\ref{tildesigma}),
with taking into account the contribution of the correlator
$\av{N_w^A(b)N^B_w(b)}-\av{N^A_w(b)}\av{N^B_w(b)}$
(see Fig.\ref{vnw11ab});
\raisebox{0mm}{+}
 - the Poisson variance: \
 $V[N_w(b)]=\av{N_w(b)}$.
}
\end{figure}

In Fig.\ref{vnwpoiab} we present the final results for the
variance of the total number of participants
in PbPb collisions at SPS energies,
%calculated with
taking into account the contribution of this correlator.
We see in particular that now the calculated
variance of the total number of participants
$V[N^A_w(b)+N^B_w(b)]$
%at $b$\,=\,$10$\,Fm
is approximately three times larger
than the Poisson one
in the impact parameter region 10-12\,Fm.

\section{Discussion and conclusions}

In the framework of the classical Glauber approach
the exact analytical expression for the variance of the number of participants
(wounded nucleons) in AA collisions at a fixed value of the impact parameter
is presented.
It's shown, that along with the optical approximation contribution
(which depends only on the total NN cross-section)
in the case of nucleus-nucleus collisions
there is the additional "contact" term contribution,
depending on the integral of the overlap of two inelastic NN cross-sections
in the impact parameter plane.
%appears.
%arises.

In the classical Glauber approach under consideration
this "contact" contribution arises
at correct taking into account
the interactions between two pairs of nucleons in colliding nuclei
(a pair in one nucleus with a pair in another).
It's found, that the interactions of higher order,
than between two pairs of nucleons,
don't contribute to the variance.
At that the expression for the mean number of participants
proved to be exact already in optical approximation,
based on taking into account only the averaged
interaction between a nucleon in one nucleus
with a nucleon in another.
The same is also valid for
the mean value and the variance of
the number of NN-collisions
in AA-interactions (see appendix \ref{ap:C}).

\begin{figure}[t]
\centerline{
\includegraphics[width=90mm,angle=-90]{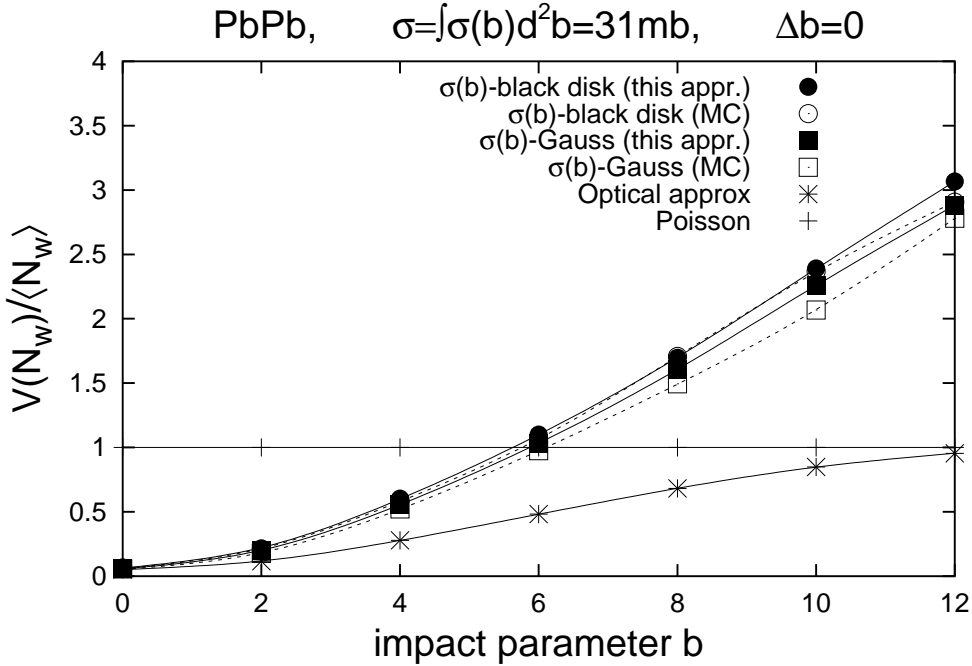}
}
\caption[dummy]{\label{vnw-nor}
The same as in Fig.\ref{vnwpoiab}, but for
the normalized variance
$V[ N_w(b)]/\av {N_w(b)}$
of the total number of wounded nucleons in colliding nuclei,
$N_w(b)\equiv N_w^A(b)+N^B_w(b)$.
}
\end{figure}

These results are obtained in
the framework of the
pure classical (probabilistic) Glauber approach \cite{Bialas76}.
However it's possible to suppose,
that in the quantum case
the one loop expression for the variance
and the "tree" expression for the mean number of participants
and NN-collisions will be exact.

Using the formulae obtained, the numerical calculations of
the variance of the participants number
in PbPb collisions at SPS energies are done.
It's demonstrated that
at intermediate values of the impact parameter
the "optical" and "contact" term contributions
are of the same order and
their sum is in a good agreement
with the results of independent MC simulations
of this process.

When calculating the variance of the total
(in both nuclei) number of participants
the correlation between the numbers of participants
in the colliding nuclei is taking into account.
The exact analytical expression for the correlator
at a fixed value of the impact parameter
is obtained.
The results of numerical calculations of
the correlator for PbPb collisions at SPS energies
show that
at intermediate and large values of
the impact parameter %$b$
its contribution to
the variance of the total number of participants
is about half of the variance in one nucleus,
again in good agreement with
%the results of
independent MC simulations.

In particular as a result it's found
that
the calculated variance of the total number of participants
in PbPb collisions at SPS energies
approximately three times larger than the Poisson one
in the impact parameter region 10-12\,Fm.
(See Figs.\ref{vnw-nor} and \ref{vncl-nor} for the normalized variance
of the number of wounded nucleons and NN-collisions.)

 Note that the good agreement of the analytical and MC
 calculations ensures the reliability both of them
 and enables to use the developed MC algorithm
 in future experimental setup motivated calculations.

The author thanks M.A.~Braun and G.A.~Feofilov for useful discussions.
The work was supported
by the grant RNP 2.2.2.2.1547 of Education Ministry of Russia
and by the RFFI grant 06-02-16115a.

%  Appendixes  XXXXXXXXXXXXXXXXXXXXXXXXXXXXXXXXXXXXXXXXXXXXXXXXXXXXX
\section*{Appendixes}
\appendix
%  App_A  XXXXXXXXXXXXXXXXXXXXXXXXXXXXXXXXXXXXXXXXXXXXXXXXXXXXXX
%\newpage
\section{Calculation of the participants variance for one nucleus}
\label{ap:A}

%%%% fig_coal %%%%%%%%%%%%%%%%%%%%%%%%%%%%%%%%%%%%%%%%%%%%%%%%%%%%
\def\figAB{\begin{figure}[t]

\unitlength=0.6mm
%\linethickness{0.5 pt}
\thicklines
\bc
%==================================
\begin{picture}(170,110)(0,-35)
%\filltype{black}
%\blacken
\put(30,45){\circle*{1}}
\put(30,45){\circle{60}}
\put(20,55){\makebox(0,0)[cc]{{\Large $A$}}}
\put(30,45){\vector(1,0){75}}
\put(40,30){\circle{10}}
\put(40,30){\circle{0.5}}
\put(40,45){\vector(0,-1){15}}
\drawline[-20](25,30)(145,30)
\put(39,39){\makebox(0,0)[rc]{$a_j$}}

%\dashline{2}(80,90)(50,80)(30,50)(10,40)
%\dashline{2}[0.5](80,80)(50,70)(30,40)(10,30)
%\dashline[-30]{2}[0.5](80,70)(50,60)(30,30)(10,20)
%\drawline(10,5)(60,10)(85,20)(90,60)(100,95)
%\drawline[-50](10,0)(65,5)(90,15)(95,55)
%\drawline[-50](10,0)(65,5)(90,15)(95,55)

\put(130,0){\circle{70}}
\put(130,0){\circle*{1}}
\put(130,-20){\makebox(0,0)[cc]{{\Large $B$}}}
\put(30,-20){\makebox(0,0)[cc]{{\large $b_{jk}=a_j-b_k+b$}}}
\put(-5,0){\line(1,0){175}}
\put(140,20){\circle{10}}
\put(140,20){\circle{0.5}}
\put(140,0){\vector(0,1){20}}
\put(90,0){\vector(0,1){45}}
\put(139,10){\makebox(0,0)[rc]{$b_k$}}
\put(89,10){\makebox(0,0)[rc]{$b$}}
\put(75,20){\vector(0,1){10}}
\put(74,25){\makebox(0,0)[rc]{$b_{jk}$}}

\drawline[-20](30,20)(150,20)

\end{picture}
%==================================
\ec
\caption[dummy]{\label{AB}
$AB$-collision.
}
\end{figure}
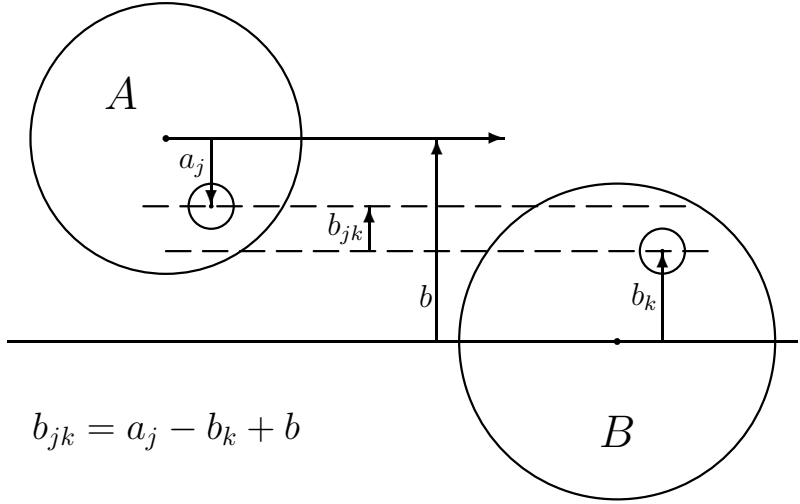}
%%%% end fig_coal %%%%%%%%%%%%%%%%%%%%%%%%%%%%%%%%%%%%%%%%%%%%%%%%%%%%

The geometry of $AB$-collision is shown in Fig.\ref{AB}.
All $a$ and $b$ are the two-dimensional vectors in the impact parameter plane.
In the framework of the classical (probabilistic) approach \cite{Bialas76}
the dimensionless $\sigma (b)$ is the interaction probability of two nucleons
at the impact parameter value~$b$:
\begin{equation}
\int \sigma (b) db = \sigma_{NN}^{in} \equiv \sigma
\label{sigb}
\end{equation}
(all integrations imply the integration over
two-dimensional vectors in the impact parameter plane).
This probabilistic interpretation means that $\sigma (b) \leq 1$,
so even in the limit $r_N\ll R_A, R_B$ we can't use approximation: $\sigma (b)=\sigma \delta(b)$.
The examples of valid approximations are as follows:
the "black disk" approximation:
\beq
\sigma (b)= \theta(r_N-|b|)\ , \hs1 \sigma=\pi r^2_N \ ,
\label{ap:black}
\eeq
the "grey disk" approximation:
\beq
\sigma (b)=\gamma \theta(r_N-|b|)\ , \hs1 \gamma<1\ , \hs1 \sigma= \gamma \pi r^2_N \ ,
\label{ap:grey}
\eeq
Gauss approximation:
\beq
\sigma (b)=C e^{-\frac{b^2}{r^2_N}}\ ,\hs1  C\leq 1\ (!)\ , \hs1 \sigma= C \pi r^2_N \ .
\label{ap:gauss}
\eeq

\figAB

$T_A$ and $T_B$ are the profile functions of the colliding nuclei $A$ and $B$.
We'll imply the factorization takes place:
\beq
T_A(a_1,...,a_A)=\prod_{i=1}^A T_A(a_i)  \ ,
\label{ap:factor}
\eeq
which is sound approximation for heavy nuclei.
Let us introduce also some shorthand notations:
\beq
\hat{d}a_i \equiv T_A(a_i) da_i \ , \hs1
\int\hat{d}a_i = \int T_A(a_i) da_i = 1 \ .
\label{short}
\eeq
\beq
\av X \equiv \avt X = \int \ol{X} \prhdb{k} \prhda{i} =
\int \ol{X} \prTdb{k} \prTda{i}
\label{avr}
\eeq
\beq
V[X] \equiv \av {X^2}-\av {X}^2
\label{var}
\eeq
Here $\ol{X}$ means average of some variate $X$ at fixed positions of
all nucleons in $A$ and $B$; $\avr{\ }{A}$ and $\avr{\ }{B}$
mean averaging over positions of these nucleons.

We introduce now the set of variates $X_1,...,X_A$ (each can
take on a value equal only to 0 or 1) by the following way:
\begin{description}
  \item[$X_j = 1$] if $j$-th nucleon of the nucleus $A$ interacts
with some nucleons of the nucleus $B$
  \item[$X_j = 0$] if $j$-th nucleon of the nucleus $A$ doesn't interact
with any nucleons of the nucleus~$B$
\end{description}
Then the number of participants (wounded nucleons) in the nucleus $A$
%at a fixed value of the impact parameter $b$
in the given event can be simply expressed through these variates:
\beq
N^A_w(b)=\sumA{j} X_j  \ .
\label{NAXj}
\eeq
So we have for the mean value:
\beq
\av {N^A_w(b)} =\sumA{j} \av{X_j} =\sumA{j} \avt{X_j}
\label{avrNA}
\eeq
and for the variance of $N^A_w(b)$:
\beq
V[ N^A_w(b)] \equiv \av{{N_w^A(b)}^2}-\av{N^A_w(b)}^2 \ , \hs1
\av{{N_w^A(b)}^2}=\av{ (\sumA{j} X_j)^2 } \ .
\label{varNA}
\eeq
Let us start our calculations from (\ref{avrNA}).
Clear that for given configuration $\{a_i\}$ and $\{b_k\}$:
\beq
\ol{X_j} =0 \cdot P(X_j=0)+1 \cdot P(X_j=1) = p_j =1-q_j \ ,
\label{olXj}
\eeq
where
\beq
P(X_j=0)\equiv q_j =  \prb{k}(1-\sig{jk}) \ ,
\label{PX0}
\eeq
\beq
P(X_j=1)\equiv p_j= 1-q_j = 1-\prb{k}(1-\sig{jk}) \ ,
\label{PX1}
\eeq
\beq
\sig{jk}\equiv\sigma(a_j-b_k+b)  \ .
\label{sigjk}
\eeq
Here $P(\,X_j=0(1)\,)$ is the probability that the variate $X_j$ will be equal to 0 or 1 correspondingly.
We have to keep in mind that $p_j$ and $q_j$ are the functions of $a_j$, $b_1$,...,$b_B$ and $b$:
\beq
q_j=q_j(a_j,\Cbk,b)\ ,
\hs 1
p_j=p_j(a_j,\Cbk,b) \ .
\label{qjpj}
\eeq

%\newpage
Straighforward calculations give:
$$
\av {N^A_w(b)} =\sumA{j} \avt{X_j}
=\sumA{j} \avAB{p_j}=\sumA{j} \avAB{1-q_j}=A-\sumA{j} \avAB{q_j}
$$
$$
\avB{q_j}=\avB{ \prb{k}(1-\sig{jk}) }= \int \prhdb{k}(1-\sig{jk})
=\prb{k} \int \hdb{k} (1-\sig{jk})
=\prb{k} (1-\int \hdb{k} \sig{jk}) =
$$
\beq
=(1-\sig{j})^B \ , \hs1
\sig{j} \equiv \int \hdb{1} \sig{j1} = \int db_1 T_B(b_1) \sigma(a_j-b_1+b) \equiv \sig{1}(a_j)
\label{sigj}
\eeq
$$
\avAB{q_j}=\avA{ (1-\sig{j})^B }= \int \prhda{i}(1-\sig{j})^B
=\int \hda{j} (1-\sig{j})^B
=\int da_j T_A(a_j) (1-\sig{j})^B =
$$
$$
= Q(b)\ , \hs1 Q(b)\equiv\int da_1 T_A(a_1) (1-\sig{1})^B ,
\hs 1
\sig{1} = \int db_1 T_B(b_1) \sigma(a_1-b_1+b) \equiv \sig{1}(a_1)
$$
\beq
\av {N^A_w(b)} =A-\sumA{j} \avAB{q_j}=A-\sumA{j} Q(b)=A-A Q(b)=A (1-Q(b))=A P(b) \ ,
\label{avNw}
\eeq
which coincides with formula (\ref{mean}) of the text.
We see that the result for the mean value of the number of participants
is the same as in an optical approximation.

%%%%%%%%%%%%%%%%%%%%%%%%
Let us now calculate the variance of $N^A_w(b)$. We start from (\ref{varNA}):
\beq
\av{{N_w^A(b)}^2}=\av{ (\sumA{j} X_j)^2 }=\av{ \sumA{j_1,j_2} \Xj{1} \Xj{2} }=
\sumA{j_1 \neq j_2} \av{ \Xj{1} \Xj{2} }+ \sumA{j} \av{X^2_j} \ .
\label{NA2}
\eeq
So we have to calculate the following two sums:
\beq
\sumA{j} \av{X^2_j} = \sumA{j} \avt{X^2_j}=\sumA{j} \avt{X_j}
\label{Xj2}
\eeq
and
\beq
\sumA{j_1 \neq j_2} \av{ \Xj{1} \Xj{2} }=\sumA{j_1 \neq j_2} \avt{ \Xj{1} \Xj{2} }
=\sumA{j_1 \neq j_2} \avr{ \avr{ \ol{\Xj{1}}\cdot \ol{\Xj{2}}  }B}A \ .
\label{Xj12}
\eeq
$$
{\rm Note\ that\ the\ last\ expression\ can't\ be\ reduced\ to\ }\hs{0.5}
\neq\sumA{j_1 \neq j_2} \avt{\Xj{1}}\cdot \avt{\Xj{2}}  \hs{0.5} (!)
$$
Just in this point the optical approximation breaks.

%%%%%%%%%%%%%%%%%%%%%%%%
Clear that
%
$$
\ol{X^2_j} =0^2 \cdot P(X_j=0)+1^2 \cdot P(X_j=1) =\ol{X_j} = p_j =1-q_j \ .
$$
Notations are the same as in (\ref{PX0})--(\ref{avNw}). And for the first sum (\ref{Xj2})
we find:
\beq
\sumA{j} \av{X^2_j} =\sumA{j} \avt{X^2_j} =\sumA{j} \avt{X_j} =\av{{N_w^A(b)}}=AP(b)=A(1-Q(b)) \ .
\label{S1}
\eeq
For the second sum (\ref{Xj12}) the straighforward calculations give:
$$
\ol{ \Xj{1} \Xj{2} }
=\ol{\Xj{1}}\cdot\ol{\Xj{2}}=\pj{1}\pj{2}=(1-\qj{1})(1-\qj{2})
=1-\qj{1}-\qj{2}+\qj{1}\qj{2} \ ,
$$
$$
\sumA{j_1 \neq j_2} \av{ \Xj{1} \Xj{2} }=
\sumA{j_1 \neq j_2} \avAB{ 1-\qj{1}-\qj{2}+\qj{1}\qj{2} }=
$$
$$
=A(A-1)-(A-1)\left(\sumA{j_1} \avAB{ \qj{1}}+\sumA{j_2}\avAB{ \qj{2}}\right)+
\sumA{j_1 \neq j_2} \avAB{ \qj{1}\qj{2} }=
$$
$$
=A(A-1)-(A-1)A(Q(b)+Q(b))+\sumA{j_1 \neq j_2} \avAB{ \qj{1}\qj{2} }=
$$
$$
=A(A-1)[1-2Q(b)+\QQ(b)] \ ,
$$
where we introduce
$$
\QQ(b)\equiv \frac{1}{A(A-1)}\sumA{j_1 \neq j_2} \avAB{ \qj{1}\qj{2} }  \ .
$$
Let us now calculate $\QQ(b)$:
$$
\avB{ \qj{1}\qj{2} }=\avB{ \prb{k_1}(1-\sig{j_1 k_1}) \prb{k_2}(1-\sig{j_2 k_2}) }
=\int \prhdb{k}(1-\sig{j_1 k})(1-\sig{j_2 k})=
$$
$$
=\prb{k} \int \hdb{k} (1-\sig{j_1 k})(1-\sig{j_2 k})
=\left( \int \hdb{1} (1-\sig{j_1 1})(1-\sig{j_2 1}) \right)^B=
$$
$$
=\left( \int \hdb{1} (1-\sig{j_1 1}-\sig{j_2 1}+\sig{j_1 1}\sig{j_2 1}) \right)^B
=\left( 1-\int \hdb{1} \sig{j_1 1}-\int \hdb{1} \sig{j_2 1}+\int \hdb{1} \sig{j_1 1}\sig{j_2 1}
\right)^B =
$$
$$
=(1-\sig{j_1}-\sig{j_2}+\ss)^B \ ,
$$
where $\sig{j_1}$ and $\sig{j_2}$ are given by (\ref{sigj}) and
\beq
\ss\equiv\int \hdb{1} \sig{j_1 1}\sig{j_2 1}
= \int db_1 T_B(b_1) \sigma(a_{j_1}-b_1+b)\sigma(a_{j_2}-b_1+b) \equiv \ssi (a_{j_1}, a_{j_2})
\label{sigj12}
\eeq
So for $\QQ(b)$ we find
$$
\QQ(b)\equiv \frac{1}{A(A-1)}\sumA{j_1 \neq j_2} \avAB{ \qj{1}\qj{2} }
= \frac{1}{A(A-1)}\sumA{j_1 \neq j_2} \avA{ (1-\sig{j_1}-\sig{j_2}+\ss)^B }
$$
$$
=\frac{1}{A(A-1)}\sumA{j_1 \neq j_2} \int \prhda{i} (1-\sig{j_1}-\sig{j_2}+\ss)^B
$$
$$
=\frac{1}{A(A-1)}\sumA{j_1 \neq j_2} \int \hda{j_1} \hda{j_2} (1-\sig{j_1}-\sig{j_2}+\ss)^B
$$
\beq
= \int da_1 da_2 T_A(a_1) T_A(a_2) (1-\sig{1}-\sig{2}+\ssi)^B \ ,
\label{ap:Q12}
\eeq
where
$$
\ssi=\int \hdb{1} \sig{1 1}\sig{2 1}
= \int db_1 T_B(b_1) \sigma(a_{1}-b_1+b)\sigma(a_{2}-b_1+b)  \equiv \ssi (a_{1}, a_{2}) \ .
$$
%
Substituting (\ref{NA2})--(\ref{ap:Q12}) into (\ref{varNA}) we find
for the variance of $N^A_w(b)$:
$$
V[ N^A_w(b)] \equiv \av{{N_w^A(b)}^2}-\av{N^A_w(b)}^2
=\sumA{j_1 \neq j_2} \av{ \Xj{1} \Xj{2} }+ \sumA{j} \av{X^2_j}-\av{N^A_w(b)}^2
$$
$$
=A(A-1)[1-2Q(b)+\QQ(b)]+A(1-Q(b))-[A(1-Q(b))]^2
$$
\beq
=AQ(b)-A^2Q^2(b)+A(A-1)\QQ(b)=AQ(b)[1-Q(b)]+A(A-1)[\QQ(b)-Q^2(b)] \ ,
\label{ap:varNA}
\eeq
which coincides with the formula (\ref{disp}) of the text.

%\newpage
The naive approach (optical approximation) implies the binomial distribution
for $N^A_w$:
%(see, for example, \cite{Wong}):
\beq
P_{opt}(N^A_w)=C^{N^A_w}_A \, P(b)^{N^A_w}\,Q(b)^{A-N^A_w},  \hs 1 P(b)=1-Q(b)
\label{opt}
\eeq
which immediately leads to
\beq
\av{N^A_w(b)}_{opt} =A P(b)
\label{opt:avr}
\eeq
and
\beq
V[ N^A_w(b)]_{opt} =AP(b)Q(b) = \av {N^A_w(b)}Q(b)\ .
\label{opt:var}
\eeq
We see that this gives true answer only for the mean value $\av {N^A_w(b)}$.
For the variance $V[ N^A_w(b)]$ the results coincide only at
$A=1$ due to explicit factor $A-1$ in (\ref{ap:varNA})
or at $B=1$ as in this case $\QQ(b)=Q^2(b)$
(i.e. for pA-collisions).
For nucleus-nucleus collisions, when $A\geq 2$ and $B\geq 2$
the naive result (\ref{opt:var}) for variance is not valid
(see text for details).

Note that for peripheral AA collisions (at large $b$),
when $P(b)$ becomes small ($P(b)\ll$1, $Q(b)\approx 1$),
the naive distribution (\ref{opt}) and the variance (\ref{opt:var}) reduce to the Poisson ones:
$V[ N^A_w(b)]_{opt} = \av {N^A_w(b)}$ (see Fig.\ref{vnwpoia}).

%  App_B  XXXXXXXXXXXXXXXXXXXXXXXXXXXXXXXXXXXXXXXXXXXXXXXXXXXXXX
%\newpage
\section{Correlation between the numbers
of wounded nucleons in colliding nuclei at fixed centrality}
\label{ap:B}

The calculations are similar to ones in the appendix \ref{ap:A}
(we use the same notations).
Allong with the set of variates $X_1,...,X_A$
we introduce in the symmetric way also the set of variates $\widetilde X_1,...,\widetilde X_B$
(each can again take on a value equal only to 0 or 1) by the following way:
\begin{description}
  \item[$\widetilde X_k = 1$] if $k$-th nucleon of the nucleus $B$ interacts
with some nucleons of the nucleus $A$
  \item[$\widetilde X_k = 0$] if $k$-th nucleon of the nucleus $B$ doesn't interact
with any nucleons of the nucleus~$A$
\end{description}
Then similarly to (\ref{NAXj}) the number of participants (wounded nucleons) in the nucleus $B$
%at a fixed value of the impact parameter $b$
in the given event can be simply expressed through these variates:
\beq
N^B_w(b)=\sumB{k} \widetilde X_k  \ .
\label{NAtXk}
\eeq

Then
\beq
\av{N_w^A(b)N^B_w(b)}=
\sumA{j}\sumB{k} \av{ X_j \widetilde X_k }=
\sumA{j}\sumB{k} \avt{X_j \widetilde X_k}
\label{XjtXk}
\eeq
and
\beq
\ol{X_j \widetilde X_k}=1\cdot1\cdot P_{jk}(1,1)+
1\cdot0\cdot P_{jk}(1,0)+0\cdot1\cdot P_{jk}(0,1)+0\cdot0\cdot P_{jk}(0,0)=
P_{jk}(1,1) \ .
\label{Pjk}
\eeq
Here $P_{jk}(0(1),0(1))$
is the probability that the variates $X_j$ and $\widetilde X_k$
will be equal to 0 or 1 correspondingly.
We have
\beq
P_{jk}(1,1)=\sigma_{jk}+(1-\sigma_{jk})\rho_{jk}\widetilde\rho_{jk}  \ ,
\label{Pjk11}
\eeq
where $\sigma_{jk}$ is the probability of the interaction
of the $j$-th nucleon of the nucleus $A$
with the $k$-th nucleon of the nucleus $B$:
\beq
\sigma_{jk}= \sigma(a_j-b_k+b) \ ,
\label{sigmajk}
\eeq
and $\rho_{jk}$ is the probability of the interaction
of the $j$-th nucleon of the nucleus $A$
with at least one nucleon of the nucleus $B$ except the $k$-th nucleon
(correspondingly $\widetilde\rho_{jk}$ is the probability of the interaction
of the $k$-th nucleon of the nucleus $B$
with at least one nucleon of the nucleus $A$ except the $j$-th nucleon):
\beq
\rho_{jk}=1-\!\!\!\!\!\!\prod^B_{k'=1(k'\neq k)} \!\!\!(1-\sigma_{jk'})\ , \hs 2
\widetilde\rho_{jk}=1-\!\!\!\!\!\!\prod^A_{j'=1(j'\neq j)} \!\!\!(1-\sigma_{j'k})  \ .
\label{rhojk}
\eeq

Combining (\ref{XjtXk})--(\ref{rhojk}) and acting as in the appendix \ref{ap:A}
we find the formulae (\ref{corr})--(\ref{tildesigma}) of the text.

%  App_C  XXXXXXXXXXXXXXXXXXXXXXXXXXXXXXXXXXXXXXXXXXXXXXXXXXXXXX
%\newpage
\section{On fluctuations of the number of collisions}
\label{ap:C}

\begin{figure}[t]
\centerline{
\includegraphics[width=90mm,angle=-90]{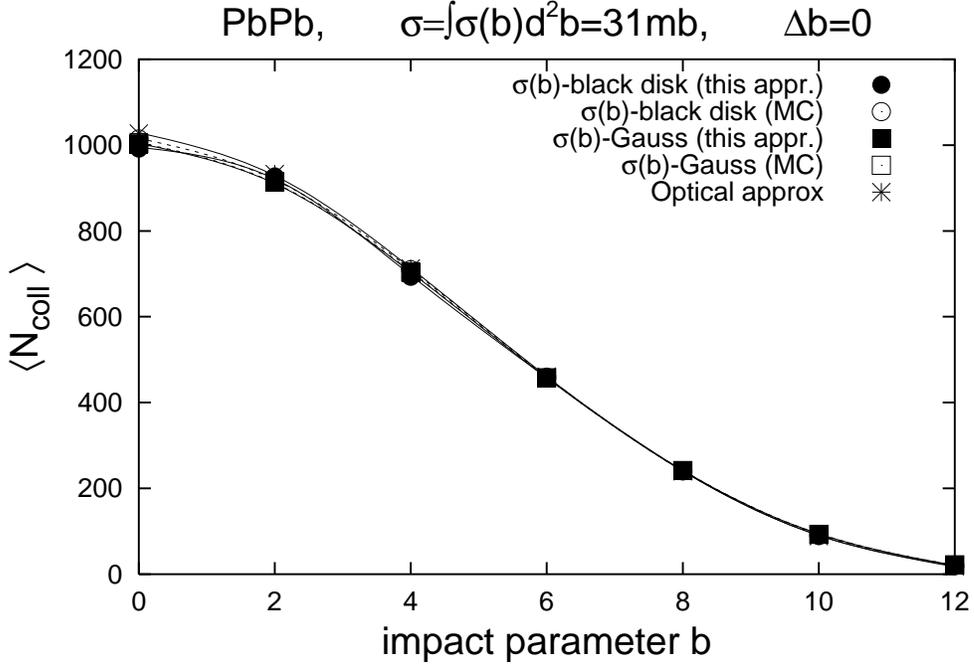}
}
\caption[dummy]{\label{ncoll}
The mean number of NN-collisions in PbPb interactions at SPS energies
calculated by formulae (\ref{avNc}) and (\ref{chi})
as a function of the impact parameter $b$; \  \
{\Large \raisebox{0mm}{${\ast}$}}
- "optical" approximation, calculated using formulae
(\ref{approxchi}) and (\ref{opt:avr:coll}).
The notations are the same as in Fig.\ref{vnwpoia}.
}
\end{figure}

In this appendix we discuss briefly the fluctuations of the number of NN-collisions
in AA-interactions at fixed value of centrality
in the framework of the approach under consideration.

To calculate the number of collisions we define the
set of $A$ variates $Y_1,...,Y_A$ (each can
take on a value from 0 to $B$) by the following way:
\begin{description}
  \item[$Y_j = 0$] if $j$-th nucleon of the nucleus $A$ doesn't interact
with any nucleons of the nucleus~$B$
  \item[$Y_j = 1$] if $j$-th nucleon of the nucleus $A$ interacts
with one nucleon of the nucleus $B$
  \item[$Y_j = 2$] if $j$-th nucleon of the nucleus $A$ interacts
with two nucleons of the nucleus $B$
   \item[...]
  \item[$Y_j = B$] if $j$-th nucleon of the nucleus $A$ interacts
with all nucleons of the nucleus $B$
\end{description}
Then $N_c(b)$ (the number of NN-collisions in the given event with impact parameter $b$)
can be expressed through these variates as follows:
\begin{equation}
N_c(b)=\sumA{j} Y_j
\label{coll}
\end{equation}
Clear that again (see appendix \ref{ap:A}):
\begin{equation}
P(Y_j=0)=P(X_j=0)= q_j =  \prb{k}(1-\sig{jk})
\label{PYj0}
\end{equation}
To calculate $P(Y_j=n)$ for $n=1,...,B$
let us introduce $\vybk$ - the random sampling from the set $\{1,...,B\}$
and $\vybkd$ - the rest after sampling. Then
\begin{equation}
P(Y_j=n) = \sum_{\vybk} \sig{jk_1}...\sig{jk_n} (1-\sig{jk_{n+1}})...(1-\sig{jk_{B}})
\label{PYjn}
\end{equation}

\begin{figure}[t]
\centerline{
\includegraphics[width=90mm,angle=-90]{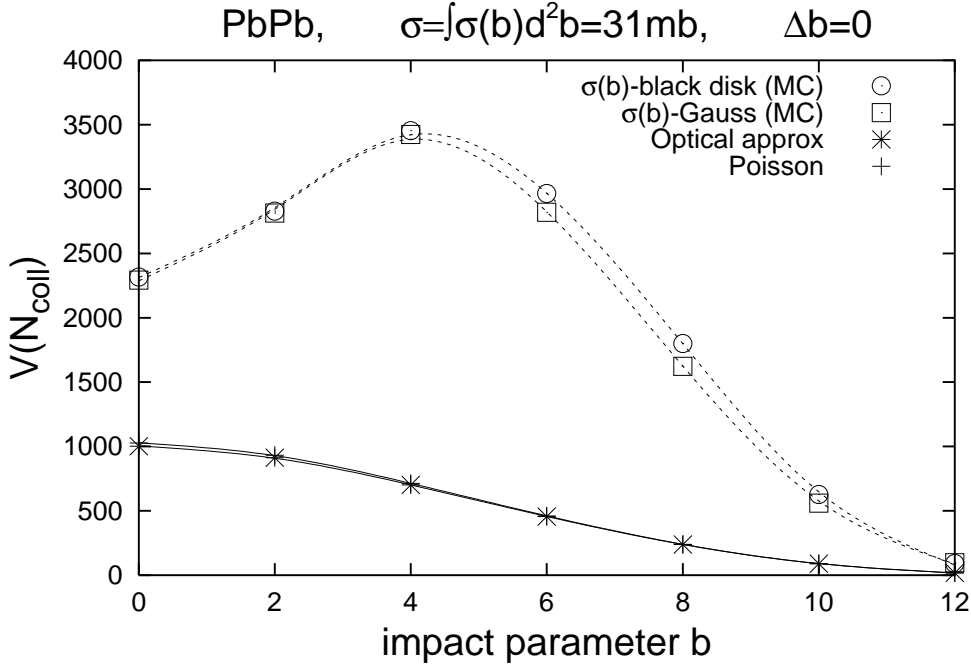}
}
\caption[dummy]{\label{vncoll}
The variance of the number of NN-collisions in PbPb interactions at SPS energies
as a function of the impact parameter $b$; \ \
{\Large \raisebox{0mm}{${\ast}$}}
- "optical" approximation, calculated using formulae
(\ref{approxchi}) and (\ref{opt:var:coll}); \ \
\raisebox{0mm}{+}
 - the Poisson variance: $V[N_c(b)]=\av{N_c(b)}$.
The notations are the same as in Fig.\ref{vnwpoia}.
}
\end{figure}

We can calculate the mean value of the number of collisions:
\begin{equation}
\av {N_c(b)} =\sumA{j} \av{Y_j} =\sumA{j} \avt{Y_j}
\label{Nc}
\end{equation}
For the given configuration $\{a_j\}$ and $\{b_k\}$ we have:
\begin{equation}
\ol{Y_j} =\sum_{n=0}^B \, n \, P(Y_j=n)
\label{olYj}
\end{equation}
and
$$
\avB{\ol{Y_j}}
=\sum_{n=0}^B \, n \, \avB{P(Y_j=n)}
=\sum_{n=0}^B \, n \, \avB{\sum_{\vybk} \sig{jk_1}...\sig{jk_n} (1-\sig{jk_{n+1}})...(1-\sig{jk_{B}})}
=
$$
$$
=\sum_{n=0}^B \, n \, \int \prhdb{k} \sum_{\vybk} \sig{jk_1}...\sig{jk_n} (1-\sig{jk_{n+1}})...(1-\sig{jk_{B}})
=
$$
\begin{equation}
=\sum_{n=0}^B \, n \, C_B^n \sig{j}^n (1-\sig{j})^{B-n}
= B\,\sig{j}  \ .
\label{avYj}
\end{equation}
We use the same notations (see (\ref{sigj})) as in appendix \ref{ap:A}:
$$
\sig{j} \equiv \int \hdb{1} \sig{j1} = \int db_1 T_B(b_1) \sigma(a_j-b_1+b)
\approx \sigma T_B(a_j+b)  \ .
$$
Finally we find:
$$
\av {N_c(b)} =\sumA{j} \av{Y_j} =\sumA{j} \avt{Y_j}
= \sumA{j} \avA{B\,\sig{j}}
= B\, \sumA{j} \avA{\sig{j}}
= B\, \sumA{j} \int \prhda{i} \sig{j}=
$$
\begin{equation}
= B\, \sumA{j} \int \hda{j} \sig{j}
= AB\, \int \hda{1} \sig{1}
\equiv AB\chi(b) \ ,
\label{avNc}
\end{equation}
where
\begin{equation}
\chi(b) \equiv \int \hda{1} \sig{1}
=\int \hda{1} \hdb{1} \sig{11}
=\int da_1 db_1 T_A(a_1) T_B(b_1) \sigma(a_1-b_1+b)
\label{chi}
\end{equation}
and at $r_N\ll R_A,R_B$
\begin{equation}
\chi(b) \approx \sigma \int da_1 T_A(a_1) T_B(a_1+b) \ ,
\label{approxchi}
\end{equation}
which coincides with the optical approximation result.
%
%\newpage

Really, assuming the binomial distribution
for $N_c(b)$ with the averaged probability $\chi(b)$ of NN-interaction:
%(see, for example, \cite{Wong}):
\beq
P_{opt}(N_c)=C^{N_c}_{AB} \ \chi(b)^{N_c}\,[1-\chi(b)]^{AB-N_c}\ ,
\label{opt:coll}
\eeq
we have
\beq
\av{N_c(b)}_{opt} =AB \chi(b)
\label{opt:avr:coll}
\eeq
and
\beq
V[ N_c(b)]_{opt} =AB\chi(b)[1-\chi(b)] = \av {N_c(b)}[1-\chi(b)] \ .
\label{opt:var:coll}
\eeq
Note that for heavy nuclei
$\chi(b)$ is small even for central collisions
($\chi(b)\sim r^2_N/R^2_A \ll $1),
and the distribution (\ref{opt:coll}) and the variance (\ref{opt:var:coll})
practically coincide with the Poisson ones:
$V[ N_c(b)]_{opt} \approx \av {N_c(b)}$.

Comparing (\ref{avNc}) and (\ref{opt:avr:coll})
we see that the optical approximation gives true answer
for the mean value $\av {N_c(b)}$.
Although we could not find the closed formula
for the variance $V[N_c(b)]$, which would be similar to formulae
(\ref{disp}), (\ref{vartot}) and (\ref{corr})
for the variance of the number of participants
$V[N^A_w(b)]$ and $V[N^A_w(b)+N^B_w(b)]$,
our calculations show that the expression for the variance of $N_c(b)$
will again differ from the optical approximation one (\ref{opt:var:coll}).
The results of our separate Monte-Carlo simulations confirm these conclusions.

\begin{figure}[t]
\centerline{
\includegraphics[width=90mm,angle=-90]{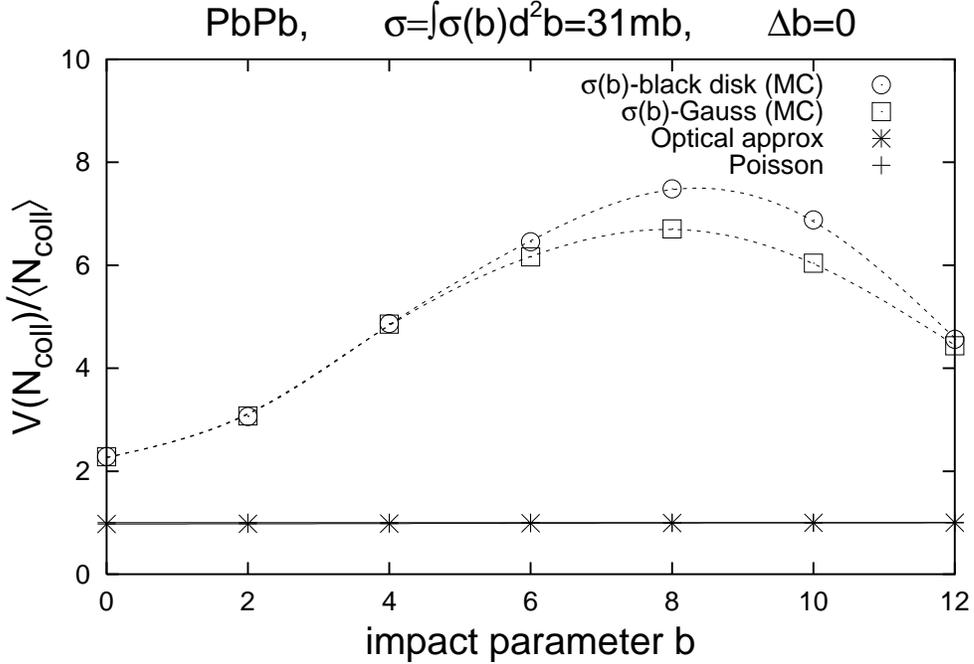}
}
\caption[dummy]{\label{vncl-nor}
The same as in Fig.\ref{vncoll}, but for
the normalized variance, $V[ N_c(b)]/\av {N_c(b)}$,
of the number of NN-collisions.
}
\end{figure}

In Figs.\ref{ncoll},\ref{vncoll} and \ref{vncl-nor} we present
the results of our calculations of the mean value $\av {N_c(b)}$
and the variance $V[ N_c(b)]$
in PbPb collisions at SPS energies
at different values of the impact parameter $b$.

%

\end{document}